# The Psychological and Physiological Part of Emotions: Multimodal Approximation for Valence Classification

Jennifer Sorinas*, Jose Manuel Ferrández and Eduardo Fernandez*, *Member, IEEE*

**Abstract**— In order to develop more precise and functional affective applications, it is necessary to achieve a balance between the psychology and the engineering applied to emotions. Signals from the central and peripheral nervous systems have been used for emotion recognition purposes, however, their operation and the relationship between them remains unknown. In this context, in the present work we have tried to approach the study of the psychobiology of both systems in order to generate a computational model for the recognition of emotions in the dimension of valence. To this end, the electroencephalography (EEG) signal, electrocardiography (ECG) signal and skin temperature of 24 subjects have been studied. Each methodology has been evaluated individually, finding characteristic patterns of positive and negative emotions in each of them. After feature selection of each methodology, the results of the classification showed that, although the classification of emotions is possible at both central and peripheral levels, the multimodal approach did not improve the results obtained through the EEG alone. In addition, differences have been observed between cerebral and physiological responses in the processing emotions by separating the sample by sex; though, the differences between men and women were only notable at the physiological level.

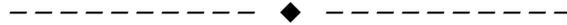

## 1 INTRODUCTION

EMOTIONS are understood as a complex set of neural and hormonal interactions that could become in affective experiences (bodily sensations); generate cognitive processes (feelings, conscious emotions); imply physiological adjustments to adapt to them; and lead to adaptive behaviors and/or decision making [1]. Emotions are an important evolutionary factor that allow survival and breeding through adaptation to the environment. However, the mechanisms of emotional processes and the modeling of human emotions are still fairly unknown. Many efforts have been made to unmask the psychobiology of emotions, since a century and a half ago Darwin proposed the first theory that tried to explain its origin [2]. Still, this is a complicated task by the fact that, even today, there is no consensus regarding the functioning, structure and classification of emotions. One of the most spread theories is the Dimensional model of emotions [3], [4], which sustains that emotions can be explained mainly by two dimensions, valence (pleasure/displeasure) and arousal (calm/excited). Depending on the level of activation and polarity of this bi-phasic dimensions, motivational systems of approach (survival and pleasure) and withdrawal (fight or flight responses) are activated with the intention of adapting behavior to an emotional stimulus [5]. Beyond the evidence founded at the brain level, this theory has increased its popularity due to the affective computing research [6]. While affective neuroscience main objective has focused on the study of the neurobiology of emotions, affective computing branch has been much more pragmatic, leaving aside the biology behind and concentrating on its recognition and classification. Nevertheless, neglecting the psychological theories of emotions has filled the affective computing studies and applications with assumptions that undermine its own credibility and effectiveness; therefore, for the development of functional affective interfaces it is necessary to contextualize engineering goals based on psychological principles [7].

In order to successful adaptation to the environment, emotions are integrated with the central nervous system (CNS) - psychological phenomena and the autonomic nervous system (ANS) - physiological part; leading to goal-directed behaviors. The fact that emotions have a psychological and a physiological part, has motivated the study of the different signals involved both in their processing and in their response to try to find patterns that allow to identify them.

At CNS level, the fMRI technique, due to its spatial resolution, is the most used for the study of the neural substrates that underlie emotions; however, the low temporal resolution, high cost and the impossibility of using it in normal life environments and situations, separate it from the picture of affective computing applications [8]. On the other hand, the temporal resolution, usability, low-cost and wireless nature of the EEG, make it the suitable technique for emotion recognition [9]. In order to

─────────────
- *J. Sorinas is with the Institute of Bioengineering of the University Miguel Hernandez, Elche 03202, Spain. And with the Department of Electronics and Computer technology, University of Cartagena, Cartagena, 30202, Spain E-mail: jennifersorinas@gmail.com*
- *JM. Ferrandez is with the Department of Electronics and Computer technology, University of Cartagena, Cartagena, 30202, Spain. E-mail: jm.ferrandez@upct.es*
- *E. Fernandez is with Institute of Bioengineering of the University Miguel Hernandez, Elche 03202, Spain. E-mail: e.fernandez@umh.es*

find the neural structures implied in the emotional process and describe the way in which they work and interact, specific frequency in the EEG spectrum, electrode location and temporal window have been studied. However, the fact that a theoretical consensus does not exist made this task complicated due to the high variability between studies and, therefore, the difficulty of comparing results. Nevertheless, asymmetry patterns across hemispheres have been observed and it is widely accepted that left hemisphere activation over frontal and prefrontal regions is linked with positive affect experience and therefore to the approach motivational system; and contrary the right hemisphere presents higher activity when processing negative affective or withdrawal stimuli [10].

Regarding the physiological part of emotions, the main structure that regulates and controls the vegetative auto-regulatory processes in order to meet behavioral demands is the ANS [11], which is closely linked with the CNS emotional part [12]; and it is believe to be involved in the generation of the physiological arousal of an emotional stimulus. Moreover, it is thought to be related with the arousal dimension but not with the valence scale [13] [14]. The ANS has two branches, the sympathetic nervous system (SNS), which becomes dominant, increasing the physiological arousal, when either psychological or physical stress is taking place; and, on the other hand, the parasympathetic nervous system (PNS) which dominates during periods of rest or safety, maintaining a low degree of physiological arousal. These systems are related with the approach and withdrawal motivational systems since they are responsible of the body response [15]; however, the degree and functionality of the process is still diffuse. The traditional view of the ANS activation stands for specific patterns of activation regarding the stimulus [16]; nevertheless, more evidences supports the undifferentiated arousal theory [17], suggesting that all emotions present the same or at least similar ANS activation pattern when high arousal stimuli take place. Several techniques have been used for emotion recognition based on bodily responses, such as heart rate, galvanic skin response, respiration, skin temperature and behavioral measures [7]. Interest in behavioral measures such as facial expressions, voice, and body language emerged because of similarities between cultures found in emotional expression [18], however these methods have a high cost and require long training time of the system and modelling of the subject [19]; therefore, physiological measures are preferred. The SNS mediated responses for negative emotions and PNS for positives have been described by several authors and measures as heart rate [20], [21], skin temperature [22], [23] and galvanic skin response [24]. However, the same problem regarding the comparison of the EEG studies apply for the ANS emotional patterns.

Each modality or physiological signal used for emotion recognition has its own pros and cons and an extensive literature behind. Calvo et al. [7] have proposed a set of factors to evaluate the effectivity of a modality to serve as a way for affective computing interfaces. First, the validity of the signal to represent the emotional process. Brain signals are preferred to physiological signals since the latter can be consciously modified and are more unspecific. Second, the reliability of the signal in real-life applications. In general, brain signals have obtained better classification results at the valence scale whereas physiological signal did for the arousal [25], [26]; suggesting that both types of signals measure different, but complementary aspects of the emotional state and therefore bringing up the idea of combining the modalities for better performances. Finally, the time resolution, cost and invasively for the user also have to be taken into account. Although EEG technology is now advancing in the development of more accessible and user-friendly devices, its complexity and discomfort is greater than that needed to measure physiological signals, which can be acquired at time through a simple bracelet [27].

On previous studies [28], [29], we have evaluated some of the technical parameters necessary for emotion recognition based on the EEG signal, but without delving deeper into its biological implications. In the present work, in addition to studying neuronal substrates underlying our computational model for the recognition of emotions based on the cerebral electrical signal; we want to study the response of the ANS and its contribution to emotion recognition on the valence scale. For this end, we have recorded the EEG, ECG and skin temperature signals of 24 subjects during stimulation using videos with positive and negative emotional content. Each biological modality has been studied individually, in a subject dependent (SD) and independent (SI) way, to finally perform a multimodal classification.

## 2 METHODS

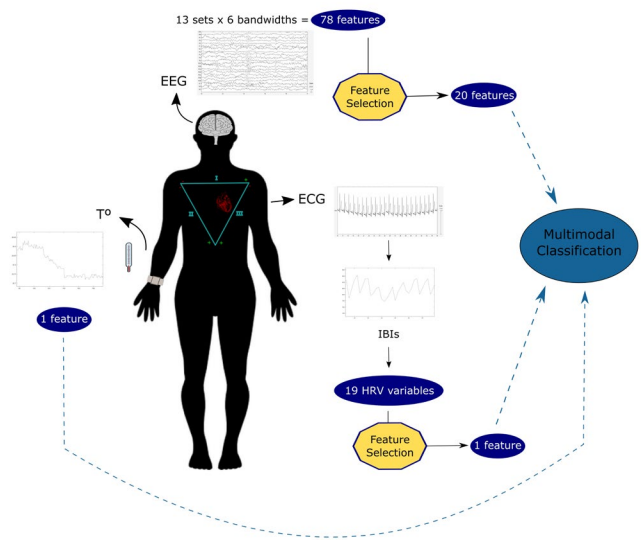

Fig. 1. Experimental setup and multimodal approach for emotion classification

### 2.1 Experimental Procedure and data analysis

A total of 24 subjects (mean age: 23.12; range: 19–37; sixteen men and eight women) were emotionally stimulated while EEG brain activity; cardiac activity and skin temperature were recorded. The stimuli consisted on 14 videos, 7 of them labelled as positive and the other 7 as negatives depending on the emotional content. The audiovisual stimuli were rated by the subjects at the valence and arousal scales, obtaining valence scores of 7.51 (s.d. 1.6) and 2.91 (s.d. 0.98) for the positive and negative categories, respectively. In the arousal scale, positive stimuli were rated with a mean arousal of 3.76 (s.d. 1.62) and 5.47 (s.d. 1.35) were rated for the negatives. The clips, with durations between 43 and 78 seconds, were selected from the internet, edited with the software Camtasia Studio 8 and presented randomly and counterbalanced to the subjects, in alternation with a 30 seconds black screen that also serves as the initial baseline. The subjects were right-handed, had no personal history of psychiatric or neurological disorders, normal or corrected vision and audition, and were not under medication during the development of the study. They also provided their written consent, supervised by the Ethics Committee of the University Miguel Hernandez.

### 2.1.1 EEG

64 Ag-AgCl electrode cap was used for the electrical brain recording according to the International 10/10 System [30]. Data were amplified and registered through a NeuroScan SynAmps EEG amplifier (Compumedics, Charlotte, NC, USA), keeping the impedance for every electrode under 25kΩ [31], and with a sampling rate of 1000Hz. Data were filtered through a high-pass and low-pass filters, 0.5Hz and 45Hz respectively, and electrodes were re-referenced to a Common Average Reference (CAR), by Curry 7 software (Compumedics, Charlotte, NC, USA). For artifact rejection corresponding to electrical non-brain origin signals as eye-blinking, heart rate and muscle activity, the Matlab toolbox EEGLAB [32] was used; artifacts were selected by means of Independent Component Analysis (ICA) [33] by means of visual inspection. Detailed data collection and pre-processing steps could be founded in previous work [28].

In our previous work [29], a set of 20 features were specified as the most informative in terms of positive and negative emotion classification; however, few could be said about the theoretical interpretation of this results. Therefore, in the present work, the study of cerebral asymmetries was performed. From the 20 frequency-location variables corresponding to Alpha, Beta1, Beta2 and Gamma frequencies at both PreFrontal (PF) left and right locations; Beta 1 and Gamma at Frontal midline (Fm); Alpha and beta1 at Central midline (Cm); Beta1 and Beta2 at Parieto-Occipital midline (POm); Gamma at both Central (C) left and right hemispheres; Gamma at Parietal (P) right; and Beta1. Beta2 and Gamma at occipital (O) right; only those which localize in one or both of the hemispheres were used to study cerebral asymmetries through the classical method [34]. The asymmetry index (AI) is calculated for the spectral power of a specific bandwidth on homologous hemispheric regions, according to the formula (1).

$$AI = \frac{right - left}{right + left} \quad (1)$$

Pairs corresponding to midline regions were not included in the analysis; therefore, 9 combinations of frequencies and locations were evaluated. The AI was calculated for every frequency-location feature in each individual subject. However, in order to look for significant differences between hemispheres and emotional categories, data from all subjects were assessed together. Mann-Whitney test was used for the statistical analysis [35] between the AI of the positive and negative emotions; and with the power spectral values themselves for the comparison between homologous hemispheric regions with either positive and negative categories. The asymmetry analysis was performed in the Matlab environment (The MathWorks Inc.).

### 2.1.2 ECG

Two electrodes were placed, one on the right side of the sternum and the other on the intercostal space between fifth and sixth rib, using lead II configuration in order to record ECG activity. The recording sampling rate was 1000Hz and, as for the EEG signal, was performed with Curry 7 technology (Compumedics, Charlotte, NC, USA). After recording, ECG data were resampled to 256Hz for further analysis. Data from subjects 1, 9, 17, 19 and 22 were not properly recorder, therefore these subjects were excluded from the analysis. Freely available stand-alone Artiifact 2.09 software [36] was used to accomplish the HRV analysis. HRV analysis was performed for every of the 24 subjects individually. A high-pass filter with the cut-off frequency at 10Hz was applied to ECG data in order to extract the interbeat interval (IBI) data from every trail. Artifact detection and elimination was carried out through cubic spline interpolation. Finally, we performed time (Mean RR, Median RR, SDNN, RMSSD, NN50, pNN50) and frequency (VLF, LF, HF, LF/HF; frequency bands 0-0.04, 0.04-0.15, 0.15-0.4, respectively. Values for the different bandwidths were obtained as a percentage, absolute values and normalized units only for the LF and HF measures) domain HRV measures. On the other hand, the Matlab environment was used to assess non-lineal methods as the Poincare plot (SD1, SD2, SD1/SD2) [37], [38], [39], [40]. See Table 1 for a detailed explanation of each HRV variable. Geometrical methods have not been included in the analysis because a minimum time period of 20 minutes' recording is necessary to ensure feasible results [41]. As expected on healthy young people, no abnormalities on the ECG signal as tachycardia, arrhythmias or bundle branch block were found.

Each of the 19 variables evaluated was z-scored and tested for normal distribution with the One-sample Kolmogorov-Smirnov test, resulting in non-normal data distribution. Mann-Whitney test was carried out to test if positive and negative data came from the same distribution at both SD and SI approximations. We have also analyzed the data separating men and women. Finally, simulated annealing optimization method was used to select the most informative features with K-nearest neighbors (KNN) with 5 neighbors and quadratic discriminant analysis (QDA) classifiers, at the SI approach.

TABLE 1
HRV VARIABLES

| | |
|---|---|
| Time Domain | **Mean RR**: mean interbeat interval <br> **SDNN**: standard deviation on NN (normal-to-normal) intervals <br> **RMSSD**: square of the root of MSSD (mean square difference of successive NN intervals) <br> **NN50**: the number of pairs of adjacent NN intervals differing by more than 50ms <br> **pNN50**: the proportion derived by dividing the NN50 by the total number of NN intervals <br> RMSSD, NN50, and pNN50 are thought to represent parasympathetically mediated HRV [41]. |
| Frequency Domain | **VLF**: very-low-frequency component (0.003-0.04Hz) <br> **LF**: low-frequency component (0.04-0.15Hz). There is controversy on whether the LF component reflects SNS activity, is a product of both SNS and PNS [42], [41] or instead it is also mainly determined by the PNS [43]. <br> **HF**: high-frequency component occurs at the frequency of adult respiration (0.15-0.4Hz), primarily reflects cardiac parasympathetic influence due to respiratory sinus arrhythmia. <br> **LF/HF ratio**: This rate is interpreted as an index of sympathovagal balance [44]. |
| Poincare Plot | **SD1**: standard deviation of the instantaneous (short-term) beat-to-beat RR interval variability. As vagal regulation over the sinus node are known to be faster than the sympathetically mediated effects, Sd1 is considered a parasympathetic index [45]. <br> **SD2**: standard deviation of the continuous long-term RR interval variability. There is evidence of both parasympathetic and sympathetic tones influenced on this index [38]. <br> **SD1/SD2 ratio**: ratio between the short and long interval variation. |

#### 2.1.3 Skin Temperature

Skin temperature signal was recorded at the right wrist of the volunteers through the ActTrust bracelet (Condor Instruments Ltda., Brazil). Due to problems during the experimentation, the data of subjects 1 and 15 could not be included in the analysis, therefore 22 subjects from the total of 24 enter into the analysis. Data were segmented in trials of 28 seconds length, corresponding to every video clip. Values more than three scaled median absolute deviations from the median were consider as outliers and replaced by the mean value. One-sample Kolmogorov-Smirnov test was used for evaluating data normal distribution, resulting as non-normal distributed data. In order to look for differences between the two categories of emotions and the neutral state, the statistical Mann-Whitney test was performed [35], in the SD and SI approximations. Gender differences were also assessed with the same statistical test.

### 2.2 Multimodal Approximation

In our previous work [29], we have proposed an EEG-based model for the classification of positive and negative emotions; in the present work we have evaluated the relevance that physiological signals could have in classification performance. In order to obtain same length segments of each type of signal, temperature data was down-sampled and ECG data was up-sampled to match EEG data. Moreover, a total of 6 subjects (numbers 1, 9, 15, 17, 19 and 22) were eliminated from the analysis due to they were missing one of the data inputs. Final classification stage was performed with the 20 variables coming from the EEG analysis, 1 variable corresponding to skin temperature, and the significant variable resulting from the ECG analysis. KNN with 5 neighbors, and QDA classifiers were selected based on former work, and applied in a SD and SI approximations. As there is evidence of gender differences in physiological signals, we also performed classification by splitting the sample by gender. Experimental procedure could be overviewed on Fig. 1.

## 3 RESULTS

### 3.1 ECG

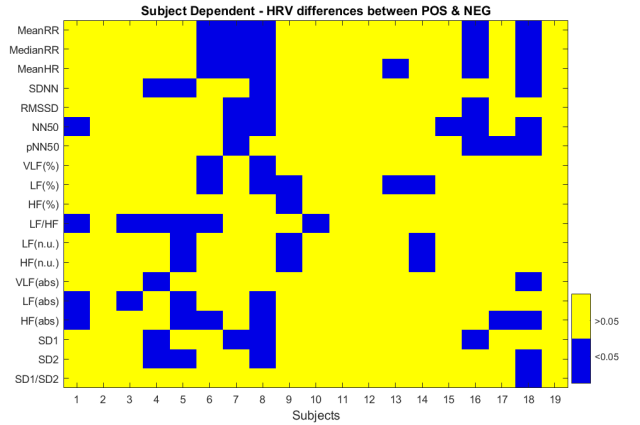

Fig. 2. P-value results for the differences between positive and negative emotions at all the HRV variables and for all subjects.

Looking at the statistical differences between positive and negative emotions at the studied HRV variables, no significant differences were found at the SI level. Contrary, significant differences appeared at the SD level for some of the variables and subjects (Fig. 2); however, no clear general patterns that allow conclusions to be drawn about the population, were found. As HRV gender differences are demonstrated in young people, we have assessed differences at the SI level for women and men. Significant differences were found between positive and negative emotions at the SD2 variable (p-value = 0.0498) for women; and at the NN50 variable (p-value = 0.0382) for men. The negative emotional category had higher values of the SD2 index than the positive category. And on the contrary, the positive emotional category presented higher values of NN50 than the negative one. This fact can be interpreted as meaning that there is greater variability between contiguous beats in the positive condition, but in the long term the variability is greater in the negative one.

Statistical analyses have revealed that making common inferences for the population based on differences in HRV measures for positive and negative emotions is not an easy task. When classifying positive and negative emotions based on all the HRV measures, performances below the value of chance were obtained for both QDA and KNN classifiers, with f1 scores equal to 0.355 ± 0.161; and 0.497 ± 0.153, respectively. After simulated annealing optimization, performance improve up to 0.57 ± 0.17 for the QDA classifier, and 0.616 ± 0.125 for KNN; through using 5 (Mean RR, RMSSD, NN50, pNN50 and VLF) and 6 (Median RR, SDNN, NN50, LF [absolute values], HF [absolute values] and SD2) HRV variables, respectively, as inputs for the sorter. Although after feature selection, classification performance improved, the selected features were not shared by the classifiers, with the exception of the NN50 index. In this context, selecting those variables that best allow discrimination between positive and negative emotions seems complicated, since the results obtained from classification, although favorable, left too much margin for error. However, since our objective was to perform a multimodal classification to evaluate the contribution of the central and peripheral nervous systems in the recognition of emotions in the valence scale, we have selected NN50 variable to form part of the final classification, since it was the only common variable in both optimization groups, although it has only shown significant differences in the men group.

### 3.2 Skin Temperature

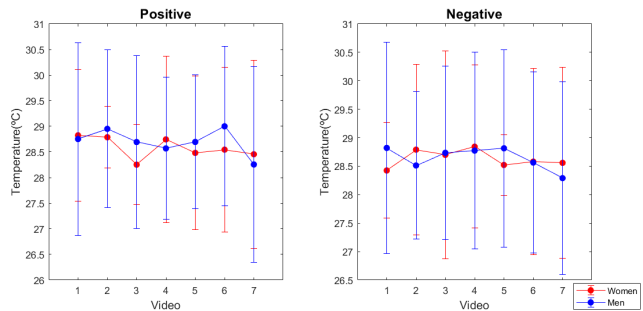

Fig. 3. Mean temperature values for every positive and negative audiovisual stimulus used. Videos are ranked from lowest to highest score on the valence scale.

Results comparing positive and negative recordings resulted in significant differences (p-value under 0.05) for all subjects in the SD approximation. Moreover, positive and negative corresponding data were also significantly different than the baseline period; except for subject 14 at both comparisons positive vs baseline and negative vs baseline. At the SI approximation, there were also significant differences between the means of the positive vs negative (p-value = 1.56e-31), positive vs baseline (p-value = 3.79e-06) and negative vs baseline (p-value = 3.79e-06) emotional groups. Mean values corresponding to positive, negative and baseline skin temperature were 28.767ºC (s.d. 1.515), 28.847ºC (s.d. 1.486) and 27.087ºC (s.d. 1.552), respectively. Temperature values for negative emotions were slightly higher than for the positive, reflecting an opposite pattern of the expected vasoconstrictor sympathetic regulation in front of adverse stimuli [23].

Looking at gender differences, skin temperature values for positive emotions showed no significant differences between women and men samples (p-value = 0.1502), but negative emotions did (p-value = 0.0017) (Fig. 3). If we disaggregate the population sample by gender, significant differences were presented in the women positive vs negative emotions (p-value = 1.7237e-42). Contrary, no differences were found for men (p-value = 0.7738).

### 3.3 EEG Asymmetries

TABLE 2
SI CLASSIFICATION F1 SCORES BASED ON OUR EEG MODEL*

|  | All population | Women | Men |
|---|---|---|---|
| QDA classifier | 0.524 (s.d. 0.082) | 0.463 (s.d. 0.082) | 0.483 (s.d. 0.076) |
| KNN classifier | 0.522 (s.d. 0.092) | 0.499 (s.d. 0.041) | 0.474 (s.d. 0.065) |

*Data coming from previous study* [29]

No differences were found between genders when classifying emotions based on EEG data (Table 2), therefore, asymmetry studies were performed with the whole population. We have studied the interhemispheric asymme-tries of the 20 frequency-location pairs, in order to have a better overview of the positive and negative emotional brain processes underlying the proposed model. This analysis has been done with either SD and SI approaches. With respect to the SD approximation, none of the AI measures had resulted as having significant differences between the positive and negative conditions in all subjects, this fact demonstrates once more the inter-subject brain variability present in the emotional processing. Therefore, in this case, performing the evaluation of the state of the frequency-location pairs would be more fruitful with a SI approach, trying to focus only on the commonalities between subjects instead of the differences.

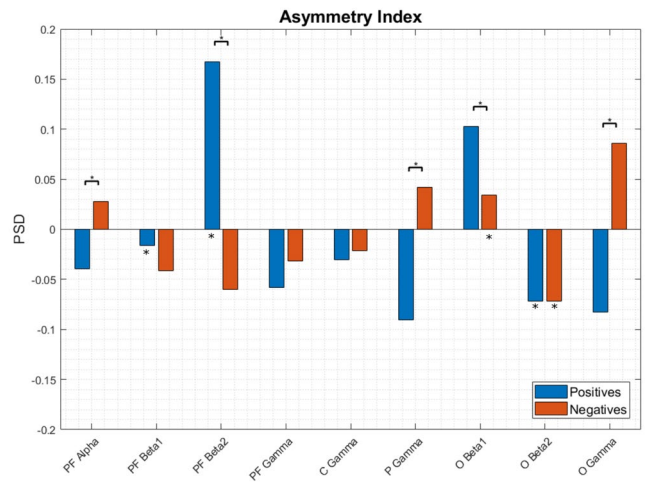

Fig. 4. AI for the 9 studied frequency-location pairs at positive and negative emotions for all subjects. The bracket with the asterisk above, which joins two bars of the histogram, indicates those pairs that have shown significant differences (p-value<0.05) between the positive-negative conditions. Asterisks on the bars indicate significant differences (p-value<0.05) between interhemispheric, intra-condition (positive-positive or negative-negative).

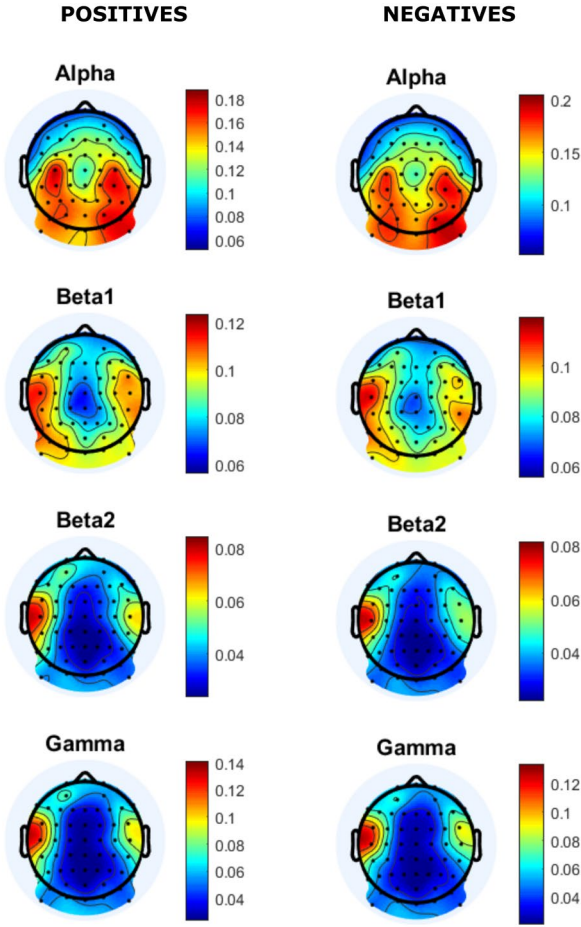

Fig. 5. Mean distribution of the spectral power of the frequencies that compose our computational model, for both emotional categories in the valence scale, in the temporal window of our model.

When comparing the AI of positive and negative emotional conditions, we have found differences in 5 of the 9 interhemispheric variables studied: PF-alpha (lateralized towards the left hemisphere at positive emotions and toward right on negative), PF-beta2 (presented the same lateralization pattern as PF-alpha), P-gamma (at both contditions lateraliaztion occurs towards the left hemisphere, however for positive emotions power spectral density is higher than for negatives), O-beta1 (positive emotions lateralized towards the right hemisphere and negatives towards the left) and O-gamma (same lateralization pattern as the presented on the PF region). If we look at the individual interhemispheric differences of each emotional category, we found significant differences in positive emotions at the PF-beta1 pair lateralizing towards the left hemisphere, the PF-beta2 lateralizing towards the left too, and O-beta2 pair lateralizing this time towards de right hemisphere. For the negativae emotions, O-beta1 and O-beta2 pairs presented significant differences, both lateralized towards the left hemisphere. Fig. 4 and 5 represent the relation between the AI and power spectral frequency at the studied frequency-location features.

### 3.4 Multimodal Approximation

After the assessment of the differences between positive and negative emotion categories at the ECG and skin temperature variables, we wanted to evaluate their contribution to the classification of emotions, assessing if it is better to take into account the central nervous system or autonomic nervous system by themselves or if it is better to use all the information, mind and body. To this end, we have performed a SD and SI classification with the KNN and QDA classifiers using different inputs. Figure 6 shows the f1 scores for the SD classification, and Figure 7 presents the SI results.

With reference to the SD classification, at both classifi-ers the T+ECG condition presented the worst performance. This result suggested that, although autonomic nervous system data per se could be used to differentiate positive and negative emotions, it is less effective than using central nervous system data alone or in combination with it. All other classifications showed similar results on both classifiers. As the performance obtained with the EEG data alone, was almost perfect, it is difficult to say if the addition of the autonomic nervous system data improves the recognition of emotions, but at least we could conclude that it does not worsen performance.

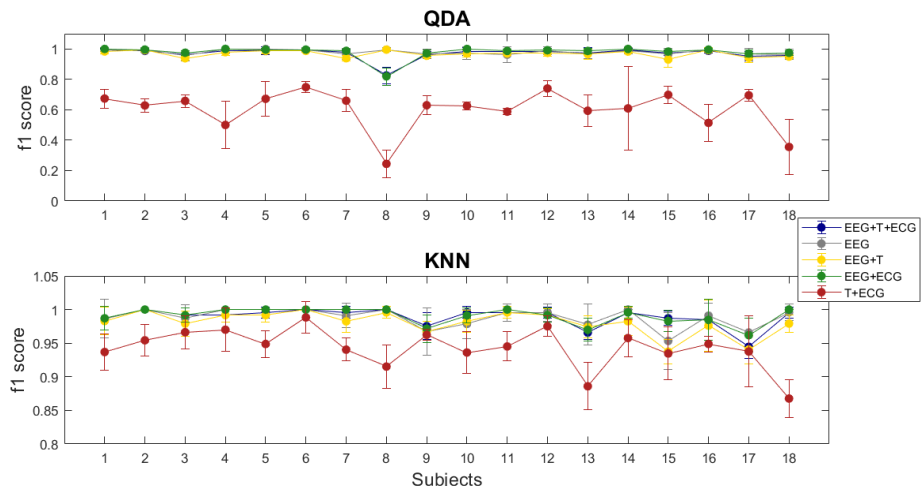

Fig. 6. F1 scores for the SD approach at the different multimodal classifications performed with the KNN and QDA classifiers

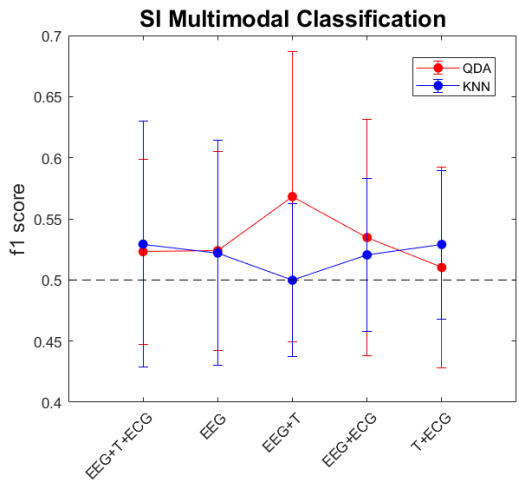

Fig. 7. F1 scores for the SI approach at the different multimodal classifications performed with the KNN and QDA classifiers. The black dotted line corresponds to value of chance.

In the case of the SI classification, we continued having the same inter-subject variability problem observed in our previous work [29]. So that, possible inferences from the results are not conclusive.

Although EEG data did not shown differences between genders, physiological signals did, therefore SI classification regarding gender was performed with all variables. The results obtained for the QDA classifier did not improve the ones obtained for the whole population (0.5231 s.d. 0.0756), the f1 scores were 0.4925 (s.d. 0.1563) for women and 0.5103 (s.d. 0.0975) for men. Contrary, classification performance improved when segregating by gender with the KNN classifier which reached f1 scores of 0.5293 (s.d. 0.1007) for the entire population, 0.5853 (s.d. 0.0558) for the women group and 0.5327 (s.d. 0.1086) for men. Although mean classification performance improved, the differences between groups were not significant.

## 4 DISCUSSION

Understanding the psychophysiology of emotional processes, i.e. the relationship between body and mind, is key to the design of effective and reliable aBCI applications. Knowing which are and how are the activation patterns of the neuronal substrates involved in the processing of emotions would allow to design more precise computational models and reduce the preparation and training times of the subjects. At the same time, being able to distinguish the emotional physiological responses and understand the performance of the ANS mechanisms responsible for them would allow a more complete emotional approach and the possibility of developing simpler and more accessible systems.

The believe that emotions are encoded in subcortical and limbic structures whereas cognition is encoded in the cortical level has been dismissed, as new evidences, coming from affective neuroscience studies, have supported the statement that emotion and cognition display or overlap along the same cortical nets [46]. In the meta-analysis conducted by Kober et al. [47], they tried to identify patterns of co-activation of brain regions and its functional organization in emotional neuroimaging studies without labeling the underlying emotions, i.e. without semantically defining the emotional category, thus overcoming the problem of lack of consensus on emotional theory. They defined six functional groups; lateral occipital or visual association group, medial posterior group, cognitive/motor group, lateral paralimbic group, medial prefrontal cortex group and the core limbic group; of which the prefrontal, occipital and central-motor cortical lobes should be highlighted. Most of these functional groups were found as relevant regions in our emotional model, thus relating the emotional process with a whole brain network, more than specific isolated areas. Moreover, another important fact revealed by the meta-analysis of Kober et al. was that all cortical structures involved in emotional processing showed co-activation with subcortical structures as the limbic system and the brainstem. The EEG only allows us to assess cortical brain activity, and therefore, it is important to note, that we are trying to classify emotions by missing an important part of the puzzle. It is thus interesting and necessary to know the body-mind relationships to have a more complete vision of the process and therefore, define emotions more accurately.

In the case of the frequency domain aspect of emotions, at positive ones, alpha and both beta frequencies seemed relevant at the left prefrontal hemisphere; and both beta1 and beta2 bandwidths increase its activation at the occipital right hemisphere. On the other hand, negative emotions presented a lateralization pattern towards the right prefrontal cortex and highlight the presence of beta1 and beta2 frequencies towards the left hemisphere over occipital regions. These results seem to point to a reverted lateralization pattern of frontal and posterior cortex when processing emotions. Our results agree with the frontal EEG asymmetry theory, described by Davidson et al. [5], [48], reflecting the activation of motivational systems of approach and withdrawal, also verified by other authors [49], [50], [51], [52]. In essence, we can conclude that there are interhemispheric differences in the processing of emotions; that this lateralization is also different depending on the emotional category; and that the processing of emotions not only falls on the prefrontal cortex, but rather there seems to be a neural network that expands along the entire cortex [53], and at all spectrum, excluding the low frequencies of the EEG [54], [53], [29]. The similarity on the scalp distribution of the different frequencies involved in the processing of positive and negative emotions suggests that, at least at the cortical level, there are no separate neural pathways for processing positive and negative emotions, but there is a network of cortical structures in charge of processing the valence, whose activity varies depending on the polarity of the emotion. Yet, a more in-depth study of the relationships between regions and emotional conditions is necessary in order to draw meaningful conclusions.

Peripheral psychophysiological reactions constitute important source of emotional information, therefore, researchers have focused on the different ANS measures. However, if the PNS and SNS are linked with the positive/approach and negative/withdrawal responses [55], respectively, is more diffuse. As for HRV measures, both our results and those of other authors, showed differences in their response to either emotional dimensions [56], [20] and discrete emotions [57], [58], [21]. These differences are present in all levels of the HRV analysis, the time domain, the frequency domain and the Poincare plot. Nevertheless, and as it is customary in the study of emotions, there is no consensus as to which are the HRV variables most representative of the emotional state. In our case, no significant differences were found in any variable with the SI approach, and although there were differences at SD level, they did not show clear patterns between individuals. Nor have different random classification results been achieved using all HRV measures to differentiate positive and negative emotions; and although after feature selection, precisions of around 60% were reached, these values are far from the percentages achieved by Guo et al. [56] or Goshvarpour et al. [20], of 71.4% and 100%, respectively. Nevertheless, when separating the sample by gender, significant differences were found between positive and negative emotions in the Poincare variable SD2 in women and in the time variable NN50 in men, corroborating the existence of gender differences in young subjects [59], [60]. The results suggest that in men the differences are more evident in the short term, responding with greater intensity to positive stimuli; and on the contrary, women respond with greater intensity to negative stimuli although the difference is observable in the longer term. This may indicate a greater readiness of men for immediate response and greater adaptability to emotional stimulation in women. The variables found as informative, SD2 and NN50, are both regulated by both components of the ANS, so conjectures about the involvement or role that each division plays in the emotional response are not possible.

In general, it has been proven that skin temperature measure could be used for positive and negative emotion discrimination [22], [23]. The accepted explanation of the role of the skin temperature in the emotional process point out to vasoconstriction responses in order to mobilize blood into the muscular system to allow reaction to an aversive stimulus. Therefore, it seems that the dichotomy between the activation of the approach and withdrawal systems apply at the skin temperature level, however, both vasoconstriction and cooling and vasodilation and warming responses, are mediated by the SNS [23], pointing to the arousal scale. Therefore, although there are different patterns of response when it comes to processing positive and negative emotions, the functional organization of the activity of the ANS components remains unclear [14]. Nevertheless, our results showed an opposite pattern of the commonly accepted response [61], obtaining higher temperatures for negative than for positive emotions. Regarding gender, although there are differences in the thermoregulation of women and men [62], [63], our results indicate that there are no gender differences at the skin blood volume regulation response when processing positive emotions, but, differences exist while negative stimuli occur; suggesting that women react more intensely to negative emotions than men.

Although the responses of the CNS and ANS systems to emotional stimuli and the relationships that exist between them are not known exactly; in both systems, independently, it is able to differentiate between positive and negative emotions. Koelstra et al. [26] used the detection of facial expressions and the EEG signal to classify emotions on the valence and arousal scales, demonstrating that classification performance improved when the two signals were combined, reaching percentages of 67.1% and 71.5%, respectively. Torres et al. [25] evaluated the combination of several biosignals for the detection of emotions on the valence and arousal scales. In the arousal scale, the best classification, 75%, was obtained after the combination of the EEG with physiological signals (heart rate, GSR, respiration and skin temperature); however, on the valence scale, the results of the combination of modalities did not improve the percentage achieved by the EEG alone, 58.75%. Our results conclude that CNS per se (0.988 f1-score classification result for the KNN classifier) is most informative than ANS data (HRV + skin temperature, 0.943 f1 score) in order to classify emotions regarding the valence dimension; and that the combination of the modalities (0.989 f1 score) does not significantly improve the results reached by the EEG alone.

Regarding the gender factor in the classification of emotions, we found differences at the level of response of the peripheral nervous system, but not at the CNS. This suggests that emotions are processed in the same way for men and women at the brain level, but the physiological response is different [64]. However, one of the main drawbacks of our study is that women and men samples were not balanced, being considerably less number of women than men, and moreover, although our sample of 24 subjects is more than acceptable for this kind of studies, when splitting it into gender, the sample size is not representative of the population for SI analysis regarding the minimum size of 15 subjects stablished for proper classifications [65]. Therefore, although our results are encouraging, it would be necessary to enlarge the sample size in order to obtain more reliable results in terms of gender differences.

At the peripheral level, it seems that there were differences between the responses to stimuli of opposing emotional valence. However, they did not seem to follow the expected pattern of 'fight or flight' or 'calm or safety' associated with the motivational systems of approach and withdrawal, that are believed to act at the level of the prefrontal EEG asymmetries. The components of the ANS are not activated in an 'all-or-none' fashion, rather each tissue is innervated differently by the sympathetic and parasympathetic pathways, which act independently of each other [14]. It is therefore difficult to attribute approaching or rejecting responses to specific components of the ANS. At this point, it is worth asking if motivational systems represent the same as the dimension of affective valence or if, on the contrary, they are different processes that do not always go hand in hand [66]. Conversely, it is probable that the arousal is influencing the physiological response, since although there were no significant differences in the arousal rating in the population, polarity existed in specific individuals.

## 5 CONCLUSION

We have assessed the activation states of the CNS, through EEG data, and ANS, through HRV and skin temperature measures, at positive and negative categories of the dimensional valence of emotion. Population differences were found at the frequency domain of electrical cortical signals showing a lateralization pattern toward the left hemisphere for the positive emotions and towards the right hemisphere for negative at anterior regions, and the inverse pattern at posterior regions. Physiological differences were also found at the skin temperature response, suggesting that valence dichotomy is also present at the peripheral level. Moreover, gender differences presented at both ANS measures but not at the CNS suggest distinct mechanisms at the central and peripheral systems and different gender predisposition. However, the multimodal classification approach did not seem to benefit emotion recognition in comparison with the existing EEG computational models. Our results bring more clarity to the debate between the theory-psychology and practice-engineering of emotions; but, more efforts are needed to finish solving the riddle of the psychobiology of emotional processes.

## ACKNOWLEDGMENT

This work was supported in part by a grant from the Ministry of Education of Spain (FPU grant AP2013/01842), the Spanish National Research Program (MAT2015-69967-C3-1), the Spanish Blind Organization (ONCE) and the Seneca Foundation - Agency of Science and Technology of the Region of Murcia.

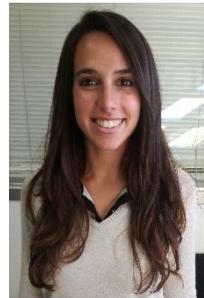

**Jennifer Sorinas** received her Degree in Biomedical Science from Universitat de Lleida, in 2013, and the Master Degree in Neuroscience from the Universidad Miguel Hernandez de Elche, Spain, in 2014, working on the development of a real-time functional near infrared system. She is currently pursuing her PhD degree in neural engineering at the same university, granted by the Government of Spain. She has performed granted research stays at Pittsburgh University – Carnegie Mellon University, USA; and at New York University, USA. Especially, she is interested in the neural substrates that codify and process emotional information.

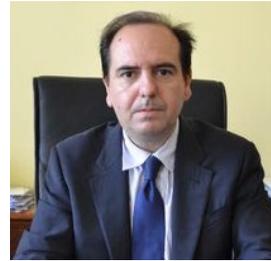

**Jose Manuel Ferrández** received the PhD degree in Informatics with honors from the Technical University of Madrid, Spain, in 1997. He made his postdoctoral stay in the Department of Neurobiology, at the University of Oldenburg, Germany. He is the director of the Electronic Design and Signal Processing Techniques Group and Vice-rector for Internationalization and Development Cooperation, at the University of Cartagena, Spain. He is an evaluator and advisor to the European Commission, in the Future Emerging Technologies Neuroinformatics program of H2020.

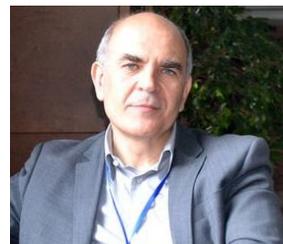

**Eduardo Fernandez** received the PhD degree from the University of Alicante, Spain, in Neuroscience with honors in 1990. He is Chairman of the Department of Histology and Anatomy in the University Miguel Hernández (Spain) and Director of the Neuroengineering and Neuroprosthesis Unit at the Bioengineering Institute at the same university. His research interest is in developing solutions to the problems raised by interfacing the human nervous system and on this basis develop a two-way direct communication with neurons and ensembles of neurons. He is actively working on the development of neuroprostheses and brain-machine interfaces.